\documentclass{IEEEtran}
\usepackage{cite}
\usepackage[english]{babel}
\usepackage[normalem]{ulem}
\usepackage{epstopdf}
\usepackage{times}
\usepackage{mathptm}
\usepackage[pdftex]{graphicx}
\usepackage{subfigure}
\usepackage{algorithmic}
\graphicspath {{./figures}}
\usepackage{balance}
\usepackage{refstyle}
\usepackage[perpage,symbol]{footmisc}
\usepackage{graphicx}
\usepackage{epstopdf}
\usepackage{graphics}
\usepackage{amsmath}
\usepackage{epsfig}
\usepackage{balance}
\usepackage[]{algorithm2e}
\usepackage{algorithmic}
\usepackage{color}
\usepackage{multirow}
\usepackage{cite}

\ifCLASSINFOpdf
 
\else
 
\fi

\begin{document}

\begin{titlepage}
\title{Towards Implementation of Robust and Low-Cost Security Primitives for Resource-Constrained IoT Devices }
\centering
\end{titlepage}

\author{Fatemeh Tehranipoor, Member IEEE\\~\IEEEmembership{San Francisco State University}, [tehranipoor@sfsu.edu]
\thanks{Fatemeh Tehranipoor is with the School of Engineering, San Francisco State University, CA, USA (e-mail: tehranipoor@sfsu.edu).}
}

\maketitle

\begin{abstract}
In recent years, due to the trend in globalization, system integrators have had to deal with integrated circuit (IC)/intellectual property (IP) counterfeiting more than ever. These counterfeit hardware issues counterfeit hardware that have driven the need for more secure chip authentication. High entropy random numbers from physical sources are a critical component in authentication and encryption processes within secure systems~\cite{sandip-kundo-2014}. Secure encryption is dependent on sources of truly random numbers for generating keys, and there is a need for an on chip random number generator to achieve adequate security. Furthermore, the Internet of Things (IoT) adopts a large number of these hardware-based security and prevention solutions in order to securely exchange data in resource efficient manner. In this work, we have developed several methodologies of hardware-based random functions in order to address the issues and enhance the security and trust of ICs: a novel DRAM-based intrinsic Physical Unclonable Function (PUF)~\cite{DRAM-maryam-ha2015} for system level security and authentication along with analysis of the impact of various environmental conditions, particularly silicon aging; a DRAM remanence based True Random Number Generation (TRNG) to produce random sequences with a very low overhead; a DRAM TRNG model using its startup value behavior for creating random bit streams; an efficient power supply noise based TRNG model for generating an infinite number of random bits which has been evaluated  as a cost effective technique; architectures and hardware security solutions for the Internet of Things (IoT) environment. Since IoT devices are heavily resource constrained, our proposed designs can alleviate the concerns of establishing trustworthy and security in an efficient and low-cost manner.
\end{abstract}


\IEEEpeerreviewmaketitle

\section{{Introduction}}
As electronic devices become ubiquitous and more interconnected, people must depend on Integrated Circuits (ICs) for the security of sensitive information. Providing this security relies on well-established primitives for key generation, data confidentiality and integrity, authentication, identification bit commitment, etc. Therefore, it is paramount for ICs to be able to perform operations and critical tasks in a low-cost yet highly secure way. Unfortunately, the conventional approaches (e.g. digital signatures, encryption) suffer from various shortcomings; they are very slow, expensive, and increasingly vulnerable to physical and side channel attacks. Hardware-based security primitives such as physically unclonable functions (PUFs) and true random number generators (TRNGs) can overcome these limitations and provide random functions in order to establish security and trustworthiness in critical application and systems. PUFs can derive secrets from the complex physical characteristics of ICs rather than storing the secrets in digital memories. PUFs can significantly increase physical security by generating volatile secrets (keys) that only exist in a digital form when an IC is powered on and operating~\cite{FPL-2017-wei}. Furthermore, a TRNG is an important security primitive used in a variety of applications including cryptographic algorithms, statistics, communication systems, simulations, etc. It is critical that a TRNG be able to produce outputs consisting of fully unpredictable and unbiased bits in a cost-effective manner. In general, these hardware security primitives should provide low-cost and efficient trustworthiness of the physical hardware platforms. One should note that while these primitives can provide advantages to ICs, there are properties and details of the design that need to be considered (e.g. power usage, overhead, heat). 

Since IoT is a rapidly emerging paradigm, the most demanding requirement for their widespread realization is security. Applying low-cost security solutions to a large scale of IoT~\cite{ryu2012survey} and even Cyber-Physical Systems (CPSs)~\cite{choo2017emerging} is possible using hardware-based security primitives such as PUFs and TRNGs. Providing a secure framework and platform for IoT systems can protect them against malicious attacks. One of the most challenging concerns for developing secure IoT devices is the resource constrained nature of these embedded systems. Security traditionally requires a great deal of resources in order to perform the computations necessary for encryption, certificate verification, third-party authentication~\cite{ICCAD-2015-nov}, etc. By implementing the previously discussed hardware security primitives, developers can easily overcome the issues of resource constrained IoT device trustworthiness and verifiability in a low-cost and efficient way.

In the following, we first present the preliminaries in Section II. In Section III, we present the proposed DRAM-based intrinsic PUFs. Section IV presents the results of our proposed techniques for DRAM-based RNG design. In section V, we will present the application of hardware security architecture for the Internet of Things. Finally, we summarize the paper in Section VI.

\section{{Preliminaries}}

\subsection{Physical Unclonable Functions}
As a means to uniquely identify chips, researchers have proposed using the random process variations that naturally occur during the manufacturing process. These effects include process variations such as the size of transistors, capacitors, resistors and other components. These are unavoidable for the most part, and must be accounted during the design and layout process. However, these random process variations can be used to our advantage if we use them to generate unique intrinsic identifiers. This is the idea behind Physically Unclonable Functions (PUFs), which was first proposed by Gassend et al. in 2002~\cite{Blaise-Gassend02}. Gassend and Pappu in 2001 developed the first silicon PUFs through the use of intrinsic process variation in deep submicron integrated circuits. They used the intrinsic process variability of silicon devices during manufacturing to produce unique, random and unclonable digital responses and called it a physically random function. They have since been called physical unclonable function to emphasize the fact that they are not repeatable. Generally speaking, PUFs should present unpredictable, robust and unclonable characteristics. PUFs are circuits that have come into prominence in the past decade and hold much promise as a hardware security primitive~\cite{TVLSI-ICAD-2017}. 

\subsection{Random Number Generations}
Cryptography and security applications make extensive use of random numbers and random bits. Random numbers are useful for a variety of purposes, such as generating data encryption keys, simulating and modeling complex phenomena, selecting random samples from larger data sets, and even for gambling. Random number generators (RNGs) are classically divided into two different types: Pseudo random number generators (PRNGs) and True random number generators (TRNGs). PRNGs are deterministic in nature, but are traditionally adequate for most applications. These type of random number generators usually require a seed (i.e. number to initialize the internal state of the generator) and the seed should be periodically changed to keep the system secure. The number sequence produced by PRNGs is random within a specific time period; meaning the method of random number generation does not provide truly random behavior. TRNGs, on the other hand, derive their randomness from a physical entropy source and provide inherently nondeterministic behavior. They are unpredictable, and are random in the entire time domain. Since TRNGs are capable of producing uncorrelated and irreproducible procedures they act as a critical component within cryptographic systems and applications. For security-centric applications the high entropy numbers from physical sources are a critical component in authentication and data encryption processes, where they are used to generate random cryptographic keys that are used to transmit data securely. Designing TRNGs around new forms of noise, one must account for certain features. Ideal TRNGs should display three essential characteristics: efficiency, non-determinism and non-periodicity. The dynamically natured variations that are induced by power supply noise exhibit the necessary characteristics.

\section{{DRAM-based Intrinsic PUFs}}
In this section, we introduce an intrinsic PUF based on dynamic random access memories (DRAM). DRAM PUFs can be used in low cost identification applications and also have several advantages over other PUFs such as large input patterns~\cite{TVLSI-2017-2016}~\cite{TDSC-April2018}. The DRAM PUF relies on the fact that the capacitor in the DRAM initializes to random values at startup. We demonstrate real DRAM PUFs and describe an experimental setup to test different operating conditions on three DRAMs to achieve the highest reliable results~\cite{May-2015-GLSVLSI}. Furthermore, we select the most stable bits use as chip ID using our enrollment algorithm. We also evaluate silicon aging effects on DRAM PUFs in details. In other words, we explore the possibility of intrinsic PUFs within Commercial Off-The-Shelf (COTS) DRAM ICs~\cite{Cryptography18}. We describe how to use the signatures to prevent modifications and uniquely identify and/or authenticate electronic devices.

\subsection{DRAM PUF Description and Properties}
PUFs intrinsic to DRAM ICs have not been explored extensively. Our primary contribution is the identification of a DRAM PUF based on startup values. We examine the effect of various operating conditions such as temperature variation, voltage variation, and aging which may influence the behavior of the DRAM PUF. Finally, we propose a selection mechanism to isolate highly stable bits within the large set of available bits in a DRAM. 

\subsubsection{DRAM PUF Advantages}
DRAMs have some unique advantages that motivated us to explore it further: \textbf{Large input pattern:}
Because of the large number of available bits in a typical DRAM, one can generate a large set of input challenges and correspondingly large output responses. This characteristic of DRAM PUF is very valuable which can make it to be distinct among all kinds of intrinsic PUFs. \textbf{Cost-effective:}
Since many computer systems have some form of DRAM on board, DRAMs can be used as an effective system-level PUF as well. It is also much cheaper than SRAM. Thus, DRAM PUFs could be a source of random but reliable data for generating board identifications (chip ID). The advantage of the DRAM PUF is based on the fact that the stand-alone DRAM already present in a System on a Chip (SoC) can be used for generating device specific signatures without requiring any additional circuitry or hardware~\cite{DRAM-maryam-ha2015}. PUFs intrinsic to DRAM ICs have not been explored extensively. Ours is one of the first works in which a DRAM has been used as a system level security Physical Unclonable Function.

\subsubsection{Startup Value Based DRAM PUF}
\begin{figure}
\centering
\includegraphics[scale=0.5]{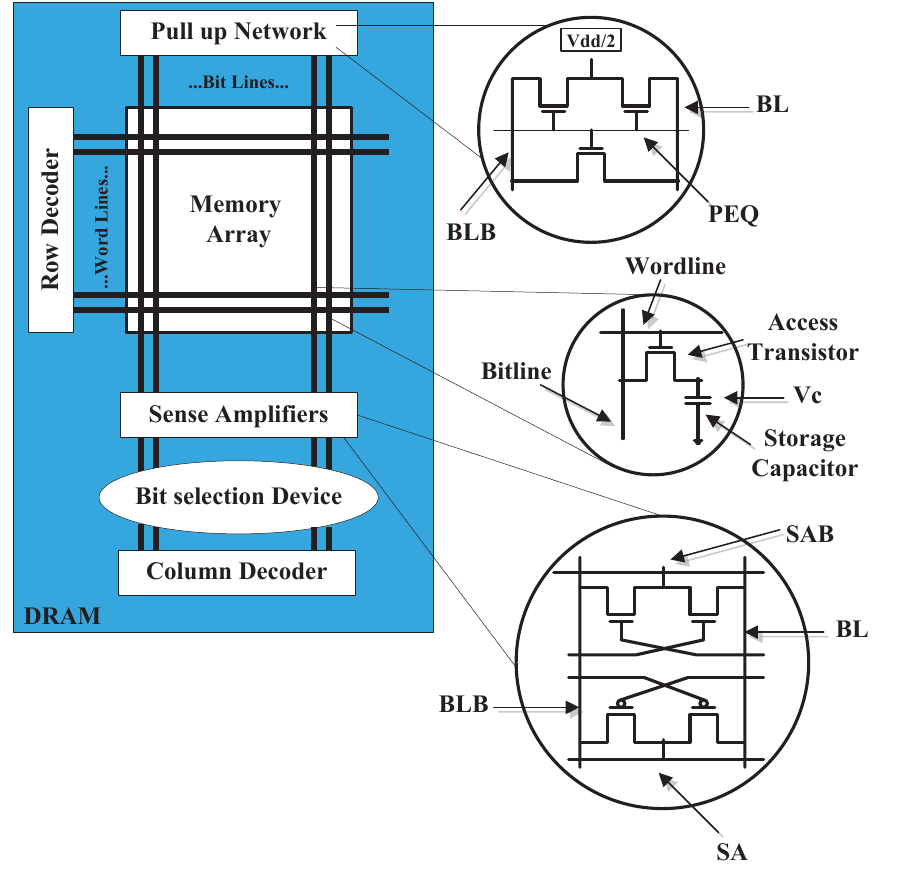}
\caption{Memory structure of a One-Transistor DRAM array.}
\label{cell2}
\end{figure}

\begin{figure}
\centering
\includegraphics[scale=0.37]{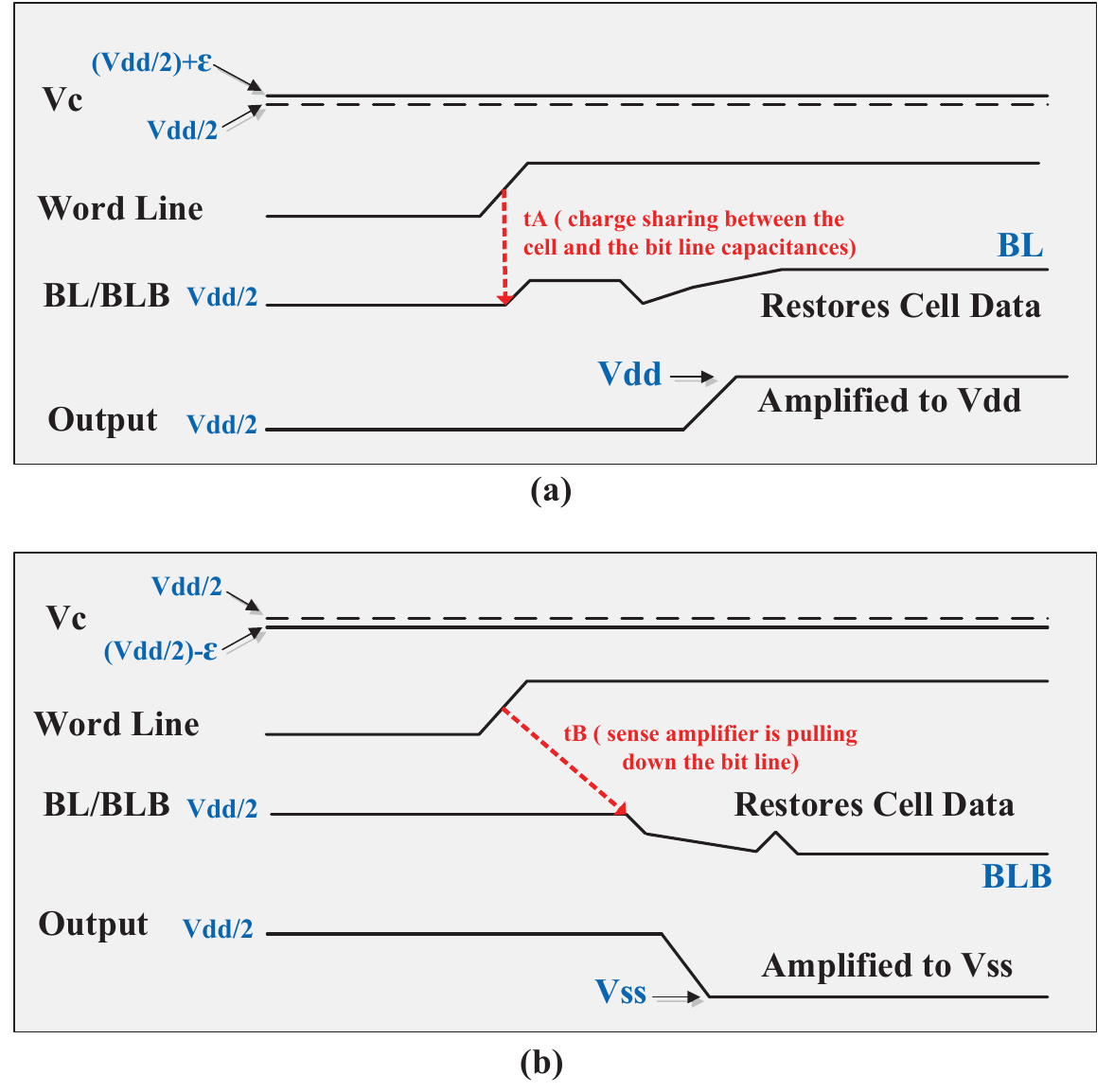} 
\caption{ Timing diagram of a DRAM read operation of an uncharged cells biased to Vdd (a) or Vss (b) due to process variations.}
\label{timingcapacitor0}
\end{figure}

In our observation of DRAM refresh and remanence properties, however, we noticed that certain DRAMs actually exhibit behavior similar to SRAMs, i.e. they have seemingly random startup values~\cite{HOST-2016-may}.  In other words, the cells do not initialize to '0' as would be expected.  Thus, as with SRAMs, these startup values provide a potential for creating a PUF. The reason for this random startup behavior can be explained by the interaction of precharge, row decoder, and column select lines when the device is powered up.  Figure~\ref{cell2} shows the structure of a typical DRAM array.  Bits are stored either by charging the storage capacitor to $V_{DD}$ or discharging it to ground.  The timing diagram of the DRAM read operation of an uncharged cells is shown in Figure~\ref{timingcapacitor0}. In order to reduce the electric field stress on the capacitor, one of the plates of the capacitor is usually biased to $\frac{V_{DD}}{2}$.  Before the reading operation, the signal to precharge the bit lines (PEQ) is disabled. In normal operation, before reading the cell, the bitlines (BL and BLB) and sensing nodes (SA and SAB) are precharged to $\frac{V_{DD}}{2}$, and when the wordline is activated, the bitlines voltage will change slightly depending on the capacitance of the storage capacitor.  This slight change is detected by the sense amplifier as a '1' (Vdd) or '0' (Vss) as shown in Figure~\ref{timingcapacitor0}. In other words, the level of BL and BLB nodes eventually reaches the operating voltage (Vdd) or  ground (Vss), respectively~\cite{kang2003cmos}. At startup, however, the storage capacitor has neither been charged to $V_{DD}$ nor discharged to ground.  Thus, at startup, the nominal voltage of each capacitor (Vc) is equal to the bias voltage $\frac{V_{DD}}{2}$ which is equal to the bitline precharge voltage.  Thus, when read, the sense amplifier is equally likely to read a '1' or '0'.  However, because of manufacturing variations, the storage capacitance of each bit will have slight differences, which leads to biasing of each bit to either a '1' or a '0'.  This behavior is what allows the startup values of the DRAM to function as a PUF~\cite{crypto-jun-2017}.

\subsection{DRAM PUFs Reliability Analysis due to Device Accelerated Aging}
Many PUFs are known to suffer from aging and lose their reliability over time - i.e. they no longer consistently return the same responses. As reliability goes down, the PUF loses its usefulness as a practical authentication device. The aging effects on a PUF are due to the degradation of transistors as a consequence of aggressive scaling in CMOS~\cite{may-2017-dram}. As technology has entered the nanometer regime, several important factors cause degradation in transistor including negative bias temperature instability (NBTI), hot carrier injection (HCI), and temperature dependent dielectric breakdown (TDDB). Since DRAM PUF behavior is primarily determined by process variations in the storage capacitor rather than variations in the transistor, it is likely that aging due to NBTI may not be an issue. In this section, we investigate the effects of aging on DRAM PUFs functionality. Our proposed approaches are evaluated on an experiment platform, the Xilinx Spartan 6 FPGA on a Digilent Atlys board. The off-chip DIP DRAMs were mounted and wired to a prototype board that has a high density serial connector during data collection. The serial connector allows the prototype board to interface with the FPGA. Then, programmed FPGA controls the test sequences applied to the DRAM and transmit the results/outputs from the DRAM chip to a computer (workstation) using a USB-UART module. All experiments are performed using our ThermoStream Burn-in System (Temptronic TP04100A ThermoStream Thermal Inducting System) to accelerate aging. Each measurement is performed at a VDD of 5V and at room temperature and consists of 144 startup state readings at different dates from Sep. 2014 to Feb. 2016. Table~\ref{stableaging}, at aged-DRAMs condition, the percentages of stable bits do not change that much especially for dram1. Furthermore, these small changes are very normal and still a large amount of data are stable during the 18-month period of time of our experiments on DRAMs. Based on our observations, our DRAMs (dram1, dram2, and dram3) which are under experiments for aging effects on PUFs functionality, are much more stable than what we even expected before starting this project, since it is a common expectation that aging has irreversible effects on devices.

\begin{table}[h]
	\centering
	\linespread{1}
	\small
	\caption{Stability of DRAMs due to Aging compare to Nominal Conditions}
	\label{stableaging}
	\begin{tabular}{c|l|l|l|lllllll}
		\cline{2-4}
		& \multicolumn{1}{c|}{dram1}             & \multicolumn{1}{c|}{dram2}             & \multicolumn{1}{c|}{dram3} & \multicolumn{1}{c}{} & \multicolumn{1}{c}{} & \multicolumn{1}{c}{} & \multicolumn{1}{c}{} &  & \multicolumn{1}{c}{} & \multicolumn{1}{c}{} \\ \cline{2-4}
		& \multicolumn{1}{c|}{stable bits (\%)} & \multicolumn{1}{c|}{stable bits (\%)} & stable bits (\%)          &                      &                      &                      &                      &  &                      &                      \\ \cline{1-4}
		\multicolumn{4}{|c|}{Pre-aging (un-aged) Condition}                                                                                                      &                      &                      &                      &                      &  &                      &                      \\ \cline{1-4}
		\multicolumn{1}{|c|}{Sep. 2014} & \multicolumn{1}{c|}{88.9\%}                  & \multicolumn{1}{c|}{91.6\%}                  &      \multicolumn{1}{c|}{89.7\%}                    &                      &                      &                      &                      &  &                      &                      \\ \cline{1-4}
		\multicolumn{4}{|c|}{Aged DRAMs}                                                                                                               &                      &                      &                      &                      &  &                      &                      \\ \cline{1-4}
		\multicolumn{1}{|c|}{Sep. 2014} &      \multicolumn{1}{c|}{87.1\%                                                                                                                                                           }                    &        \multicolumn{1}{c|}{90.1\%                                                                                                                                                           }                                &     \multicolumn{1}{c|}{80.2\%}     &                      &                      &                      &                      &  &                      &                      \\ \cline{1-4}
		\multicolumn{1}{|c|}{Feb. 2015} &     \multicolumn{1}{c|}{86.4\%}                                   &        \multicolumn{1}{c|}{85.1\%}              &       \multicolumn{1}{c|}{83.1\%}            &                      &                      &                      &                      &  &                      &                      \\ \cline{1-4}
		\multicolumn{1}{|c|}{Mar. 2015} &      \multicolumn{1}{c|}{90.2\%}                                                                    &                 \multicolumn{1}{c|}{83.2\%}                                                                                          &                  \multicolumn{1}{c|}{78.0\%}                                                                             &                      &                      &                      &                      &  &                      &                      \\ \cline{1-4}
		\multicolumn{1}{|c|}{Apr. 2015} &       \multicolumn{1}{c|}{85.8\%}                                                                                                    &                 \multicolumn{1}{c|}{82.4\%}                                                                                          &                \multicolumn{1}{c|}{76.6\%}                                                                               &                      &                      &                      &                      &  &                      &                      \\ \cline{1-4}
		\multicolumn{1}{|c|}{Jul. 2015} &      \multicolumn{1}{c|}{87.3\%}                                                                                                     &               \multicolumn{1}{c|}{81.7\%}                                                                                            &                 \multicolumn{1}{c|}{81.3\%}                                                                              &                      &                      &                      &                      &  &                      &                      \\ \cline{1-4}
		\multicolumn{1}{|c|}{Aug. 2015} &     \multicolumn{1}{c|}{86.7\%}                                                                                                      &               \multicolumn{1}{c|}{81.2\%}                                                                                            &                  \multicolumn{1}{c|}{80.1\%}                                                                             &                      &                      &                      &                      &  &                      &                      \\ \cline{1-4}
		\multicolumn{1}{|c|}{Jan. 2016} &     \multicolumn{1}{c|}{86.2\%}                                                                                                      &               \multicolumn{1}{c|}{82.7\%}                                                                                            &                  \multicolumn{1}{c|}{82.4\%}                                                                             &                      &                      &                      &                      &  &                      &                      \\ \cline{1-4}
		\multicolumn{1}{|c|}{Feb. 2016} &     \multicolumn{1}{c|}{87.5\%}                                                                                                      &               \multicolumn{1}{c|}{84.6\%}                                                                                            &                  \multicolumn{1}{c|}{81.9\%}                                                                             &                      &                      &                      &                      &  &                      &                      \\ \cline{1-4}
	\end{tabular}
\end{table}


\section{DRAM-based RNG Design}

\subsection{Hardware TRNG using DRAM Remanence Effects}
\begin{figure}
	\centering
	\includegraphics[scale=0.57]{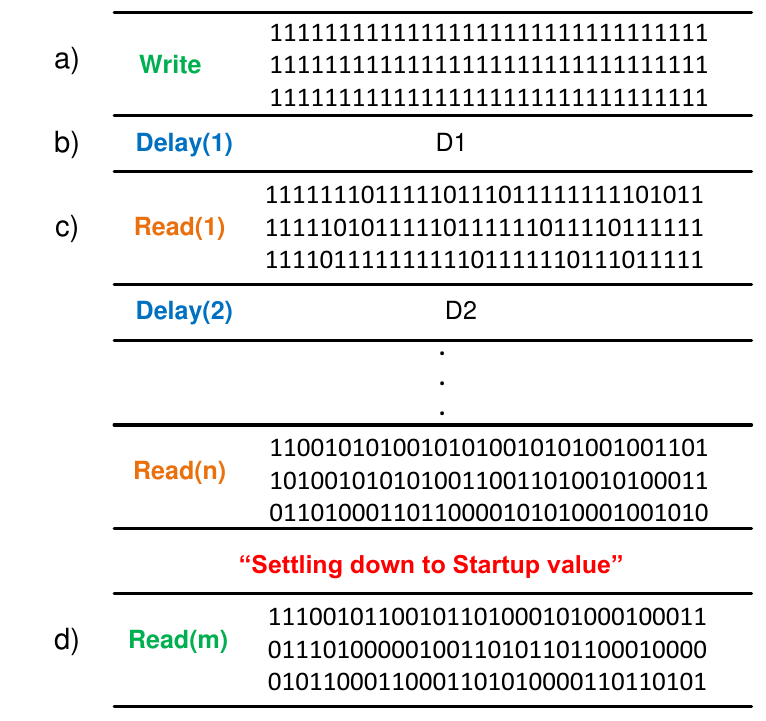}
	\caption{Real data snippets that illustrate the DDR2 SDRAM operations: a) Write. b) Delay(1). c) Read(1). d) Read(m).}
	\label{figure-snipped2}
\end{figure}

Here, we describe our methodology of using the DRAM remanence effect to propose a new TRNG model. Data remanence is the residual information that remains on a storage medium even after erasure (data clearing) or powering off the device. We start with a brief review of a typical DRAM architecture. A DRAM memory cell uses a single transistor and a capacitor to store a bit of data. Cell information (voltage) is degraded mostly due to a junction leakage current at the storage node. Therefore, the cell data must be read and rewritten periodically even when memory arrays are not accessed. Essentially, the DRAM controller must refresh each cell voltage before it decays to the point where the bit information gets lost. Normally, the refresh rate is so high that each cell gets refreshed several times per second. The processes of extracting random bits from DRAMs, while considering the remanence effect and startup behavior of DRAMs is briefly laid out as follows. As shown in Figure~\ref{figure-snipped2}, the process is first we write the value '1' to all cells of the available memory; this can be seen as step 'a)' in Figure~\ref{figure-snipped2}. After the write operation, a delay function (step b) has been applied to turn OFF the DRAM for certain amount of time, which is in milliseconds, and then turn the DRAM back on after a specific delay time. In the next step, the entire 1 Mbit of data are read (step c) and stored.

\subsubsection{NIST Statistical Test Suite Results}
The NIST statistical test suite is used to evaluate the ``randomness" of the bit strings produced by DRAM cells. Based on this test, we can determine whether a data set has a recognizable pattern or the process that has been generated is significantly random. The NIST Test Suite (NTS) is a statistical package consisting of different types of tests to evaluate the randomness of binary sequences. Each statistical test is employed to calculate a P-value that shows the randomness of the given sequences based on that test. If a P-value for a test is determined to be equal to 1, then the sequence appears to have perfect randomness. A p-value $>= 0.01$ (normally 1\%) would means that sequence would be considered to be random with a confidence of 99\%~\cite{rukhin2001statistical}. To evaluate our collected data from DRAM, we apply NIST tests to the generated bitstreams.  Our results sows that all the NIST tests p-value are greater than 0.01, this indicates that the measurements pass the requirements for randomness.

\subsection{DRAM Startup Value based RNG}
In this section, we demonstrated that DRAMs surprisingly have startup values - i.e. non-zero values when the DRAM is powered on. While older DRAMs may show potential as a PUF, our work with modern DDR DRAMs show that they not satisfy criteria to serve as a PUF. Specifically, these startup value patterns were neither random or reliable. However, we will show that DDR DRAMs can still be used to generate random numbers. Using a variety of correction mechanisms, we are able to improve the randomness of the numbers such that they pass the NIST tests. 

\begin{figure}
	\centering
	\includegraphics[scale=0.2]{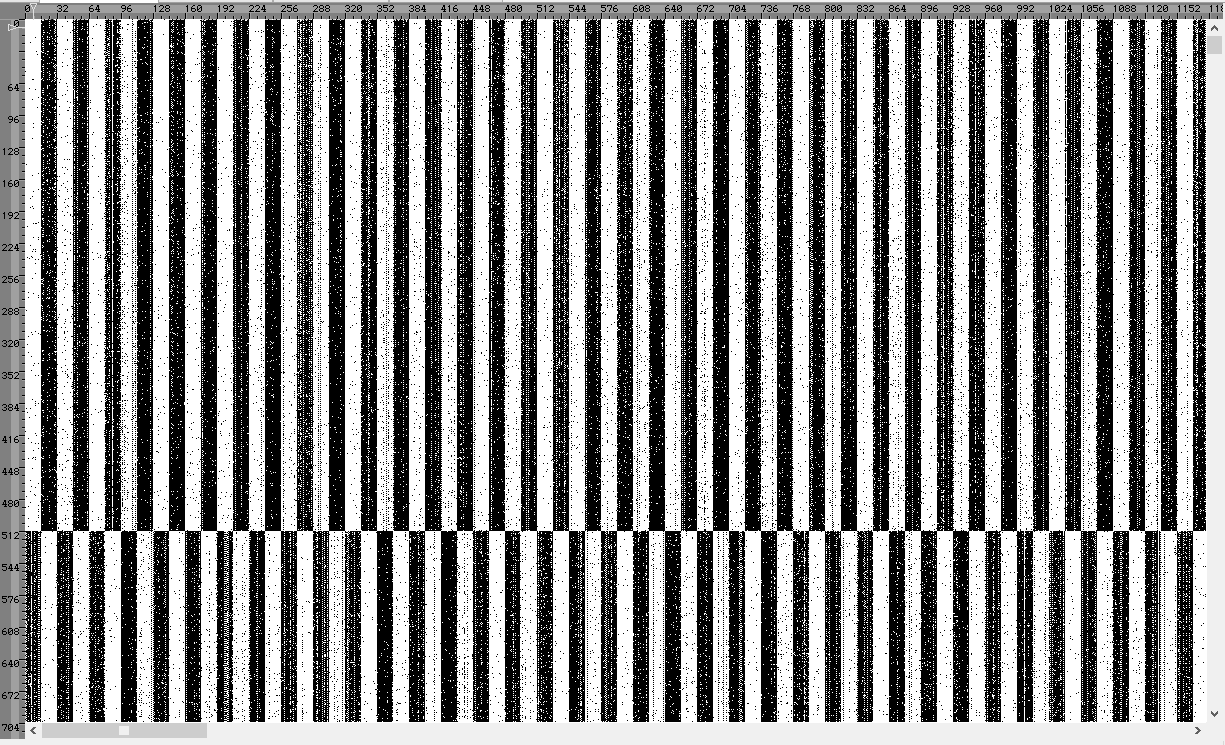}
	\caption{Small section of the bitmap from DRAM for trial 3.}
	\label{figure2-DRAM}
\end{figure}

\begin{figure}
	\centering
	\includegraphics[scale=0.2]{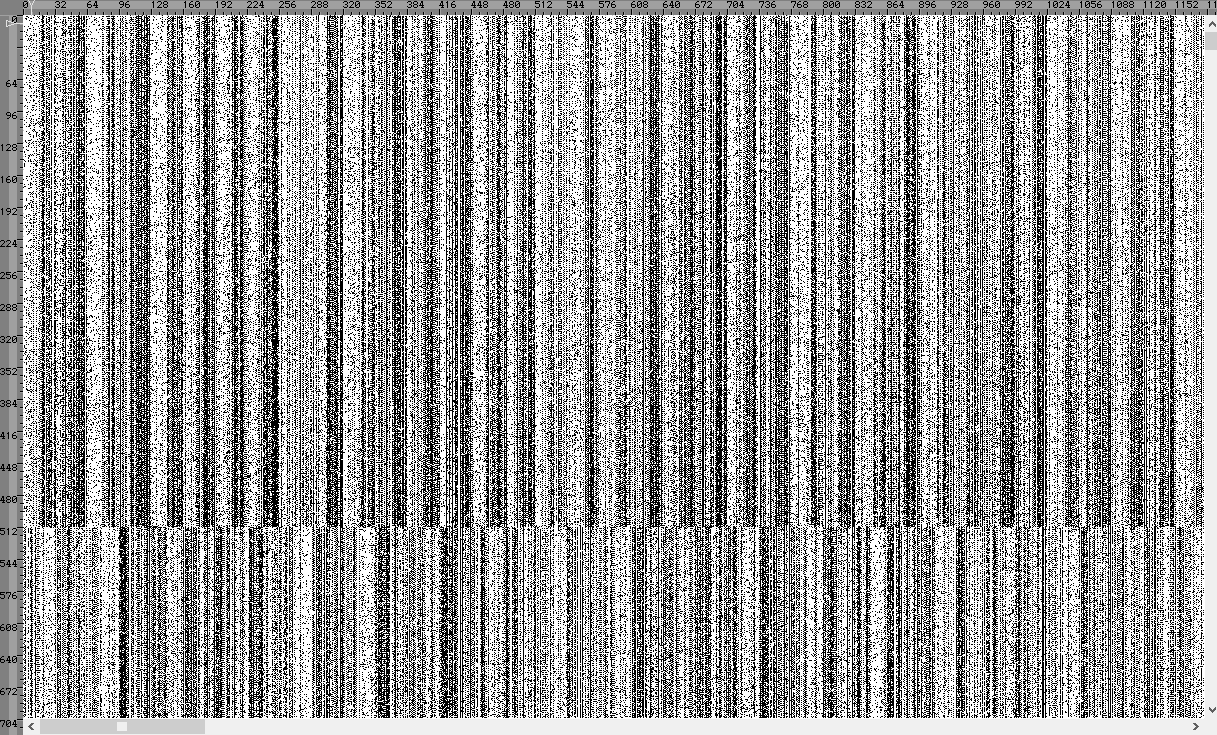}
	\caption{Small section of the bitmap from DRAM for trial 6.}
	\label{figure3-DRAM}
\end{figure}

The use of on-board DRAMs for PUFs has tremendous promise because of the large memory space, but it also contains large potential drawbacks. In our experiments, we found the startup values show a clear bias that reflects the architecture of the DRAM. In addition, not all trials produce adequate results. At times, the startup values are completely non-random yielding no valid data to be used for a PUF. More research is needed to see what is exactly causing the DRAM to start-up to these modes and if it can be avoided. Thus multiple trials will be needed to ensure that the bits are behaving correctly upon startup. We show a graphical representation of a sample bitmap from various trials in Figures~\ref{figure2-DRAM}, and~\ref{figure3-DRAM}. Each row in the Figure represents 8192 bits where white is 0 and black is 1. Across the whole device, clear patterns could be seen when mapping the data to a bit map. The architecture of the DRAM is what is likely strongly influences this style of DRAM. These patterns can be observed across multiple trials. This particular DRAM alternate between ones€™ and zeros€™ every 16 bits consistent with the DRAM 16-bit width. Every four megabits, the DRAM has a section of 32 bits before returning to its pattern. Figure~\ref{figure2-DRAM} shows a small subsection of the bitmap of one of the trials on the DRAM. The pattern described above is heavily noticeable in the figure. However, not every trial gives these results. Some of the trials had far less stable bits and the patterns were less obvious. In Figure~\ref{figure3-DRAM}, you can see the a much larger percentage of bits that don't follow the pattern compared to Figure~\ref{figure2-DRAM}. In Figure~\ref{figure3-DRAM} despite the higher variance it bits, the original pattern is still easily discernible to the human eye despite the significantly larger number of bits that do not follow the pattern. Of the six trials, Figure~\ref{figure2-DRAM} appeared to follow the pattern the best while the trial from Figure~\ref{figure3-DRAM} followed the pattern the least. The other trials were anywhere in-between. Another interesting result comes when one looks at how the pattern starts. In Figures~\ref{figure2-DRAM} and~\ref{figure3-DRAM}, one can observe that the pattern starts with 16 bits of zeros first before alternating to 16 bits of ones. However, this is not always the case. In conclusion, in our tests using a newer DDR2 DRAM, we demonstrate that the startup values are not suitable for device authentication but they do have use for creating random keys~\cite{charlie-August-2017}. 

\subsection{Power Supply Noise based TRNG}
This section introduces the construction of True Random Numbers Generators (TRNGs) using variations in power supplies. We demonstrate that power supply line outputs do not have a constant voltage and the variations in voltage follow a normal distribution. These variations can be used to create truly random bits that demonstrates a high entropy rate based on the results obtained from the NIST Statistical Test Suite. In order to quantify the impact of variations on the input signal of a circuit, we analyze the impact of such variations using Monte- Carlo simulations as well as an actual implementation. Results were obtained for evaluating the accuracy and randomness of the data gathered from our proposed circuit. A detailed analysis of the effect of variations of different power supplies is also presented with observations on their usefulness as a TRNG. The key advantage of our power supply variation based TRNG is its simplicity of implementation. For this work, we have considered 5 different power supplies; Bench power supply, USB, Computer power source and DC power supply~\cite{VLSID-Jan-2017}.

\subsubsection{Dynamic Voltage Feedback Tuning (DVFT) Design}
\begin{figure}
\begin{center}
\includegraphics[scale=0.44]{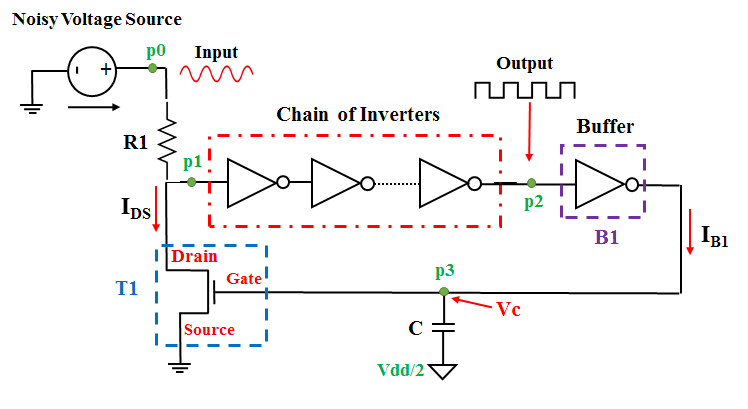}
\caption{DVFT circuit model for the proposed TRNG.}
\label{Figure-DVFT}
\end{center}
\end{figure}

To produce robust design of our TRNG circuit, we added a few new elements to our primary design (inverter chains - even and odd numbers) to allow for self-calibrating design and improved functionality.  Our RNG circuit design removes the need for manually finding the ideal voltage and moreover, keeping it steady during operation, allowing for the system to arrive at this voltage automatically using dynamic voltage tuning. Our design solution is a feedback circuit that calibrates the voltage until we have reached the optimal setting, as shown in Figure~\ref{Figure-DVFT}.  This TRNG circuit is the Dynamic Voltage Feedback Tuning (DVFT)~\cite{TVLSI-2018-2804258} which includes a buffer (B1), a precharged capacitor (C), and a transistor (T1).   The buffer, implemented through the use of an inverter, is used to isolate the TRNG output from the feedback circuit.  The capacitor serves as an integral controller essentially summing up the past history of 1's and 0's.  Finally, the transistor serves as the mechanism to tune the voltage by varying the effective resistance in the voltage divider. The advantage of this self-adjusting system is that it can maintain itself within operation range.  In other words, the system is flexible enough to recover from fluctuation in voltage coming from the power supply by attempting to prevent potential mistakes and malfunctions from occurring.  As shown in Figure ~\ref{Figure-DVFT} it is a very simple design that barely adds overhead to the entire TRNG system.


\subsubsection{Experimental Setup and Results}
In order to fully optimize our system, selecting the best hardware and software options by balancing performance and user friendly was absolutely necessary. The first step of implementing the TRNG was to determine appropriate hardware to run Linux on. Initial research suggested that a BeagleBone, a Xilinx Nexys 4 DRR FPGA, and a Raspberry Pi would be suitable to implement the TRNG on. Hardware choices were compared by the maximum bit-rate an input pin could be read by software, the amount of time required to get a Linux based operating system running on the hardware, the cost to acquire the hardware, and the time required to obtain the hardware. 

%

\section{Hardware Security Architecture for the Internet of Things}
The Internet has drastically changed the way we live, moving interactions between people at a virtual level in several contexts spanning from professional life to social relationships. The origin of the Internet of Things (IoT) can be traced back to the development of the Internet (interconnected network of computer networks)~\cite{Books-chapter18}. IoT is a novel paradigm that is rapidly gaining ground in the scenario of modern wireless telecommunications. Another basic property of these things is push button connectivity to the Internet or peer devices. IoT is comprised of a number of technological protocols that aim to make up connections from one object to other things and databases. The reason that one would to connect IoT devices is to exchange information and monitor data which needs to be secured. The first place where trust needs to be established in at the hardware level which is the platform that any software runs on top of. IoT still has many challenging issues that need to be addressed and both technological as well as social knots that have to be untied before the IoT idea becomes more widely accepted. In this paper, we proposed a solution to the results of our investigation of the infrastructure of IoT devices in their application to authentication within the healthcare domain~\cite{ICCE-2018-jan}~\cite{CODES-2016-oct}. Through our solution one would know that the information produced from an IoT is trustworthy, that the individuals accessing information can be properly identified, and that the exchange of sensitive biological signals across a network is secure. Furthermore, we showed that the hardware-based solutions are low-cost and feasibly implementable on existing resource constrained IoT systems (sensors, wearables, smart devices)~\cite{embc-BHI-paul}.

\section{Summary}
In this paper, we summarized our works for the implementation of robust and low-cost security primitives for resource-constrained IoT devices. First, we proposed a novel dynamic memory based PUF (DRAM PUF) for the authentication of electronic hardware systems. The DRAM PUF relies on the fact that the capacitor in the DRAM initializes to random values at startup time. Most PUF designs require custom circuits to convert unique analog characteristics into digital bits but with using our method, no extra circuitry is required to achieve a reliable 128 bits PUF. Our results showed that the proposed DRAM PUF provides a large number of input patterns (challenges) when compared to other memory-based PUFs circuits such as Static RAM PUFs. Our DRAM PUFs provided highly unique PUFs with a 0.4937 average inter-die hamming distance. We also proposed an enrollment algorithm to achieve highly reliable results to generate PUF Identifications for system level security. This algorithm has been validated on real DRAMs with an experimental setup to test different operating conditions. 

Second, we presented a robust hardware TRNG based on the Dynamic RAM (DRAM) remanence effect, which is a condition whereby information remains in a DRAM even after powering it down. The advantage of our hardware TRNG is that it forms from existing components with no extra circuitry. We assessed and tested the randomness of our proposed hardware TRNG by applying the NIST Statistical Test which indicates the unpredictability and non-repeatability of our data. Given its strong NIST results, we believe that there is a potential for immediate cryptographic applications. 

Third, we considered the design and implementation of a low-cost and lightweight TRNG. In the interest of being thorough, we examined six different power supplies in order to verify the non-cyclostationary behavior of the voltage sources. Our novel TRNG model is based on power supply variations (noise behavior) and self-adjusting operation. The benefits of this novel design are that: it is simple and easy to implement, there is little to no additional cost required to incorporate the TRNG into existing circuitry, and the addition of Dynamic Voltage Feedback Tuning (which we call DVFT) allows for feasibility and robustness of our model. The cumulative affect of these benefits is the practicality of the entire power-supply noise based TRNG system. We then validate the results of our theoretical model and experimental setup to show that there is a high entropy rate based on the findings from the NIST Statistical Test Suite. Based on our observations and results, our DVFT power-supply noise based TRNG model has the potential to be used in critical applications while also having the advantage of simplicity and practicality. 

In summary, we developed a series of designs and architectures based on hardware random function to tackle the issues and vulnerabilities to hardware objects, in order to protect them from malicious attacks, counterfeiting, reverse engineering, etc. Our hardware-based random functions security primitives provide low-cost, lightweight, efficient, and secure hardware platforms for the embedded systems.

\end{document}